\documentclass[11pt,preprint]{aastex}

\newcommand{\VEV}[1]{\langle#1\rangle}

\shorttitle{Foreground subtraction using WI-FIT}

\usepackage{epsfig}

\begin{document}

\title{Foreground Subtraction of Cosmic Microwave Background Maps using WI-FIT (Wavelet based hIgh resolution Fitting of Internal Templates)}

\author{F. K. Hansen}
\affil{Institute of Theoretical Astrophysics, University of Oslo,\\ P.O. Box 1029 Blindern, N-0315 Oslo, Norway; \\Centre of Mathematics for Applications, University of Oslo, \\P.O. Box 1053 Blindern, N-0316 Oslo }
\email{frodekh@astro.uio.no}

\author{A. J. Banday}
\affil{Max Planck Institut fur Astrophysik, Karl Schwarzschild-Str. 1, Postfach 1317, D-85741 Garching bei Munchen, Germany}
\email{banday@MPA-Garching.MPG.DE}

\author{H. K. Eriksen}
\affil{Institute of Theoretical Astrophysics, University of Oslo, \\P. O. Box 1029 Blindern, N-0315 Oslo, Norway; \\Centre of Mathematics for Applications, University of Oslo, \\P.O. Box 1053 Blindern, N-0316 Oslo; \\Jet Propulsion Laboratory, M/S 169/327, 4800 Oak Grove Drive, Pasadena CA 91109} 
\email{h.k.k.eriksen@astro.uio.no}

\author{K. M. G\'orski}
\affil{Jet Propulsion Laboratory, M/S 169/327, 4800 Oak Grove Drive, Pasadena CA 91109; \\Warsaw University Observatory, Aleje Ujazdowskie 4, 00-478 Warszawa, Poland; \\California Institute of Technology, Pasadena CA 91125}
\email{Krzysztof.M.Gorski@jpl.nasa.gov}

\and

\author{P. B. Lilje}
\affil{Institute of Theoretical Astrophysics, University of Oslo,\\ P. O. Box 1029 Blindern, N-0315 Oslo, Norway; \\Centre of Mathematics for Applications, University of Oslo, P.O. Box 1053 Blindern, N-0316 Oslo}
\email{lilje@astro.uio.no}

\begin{abstract}

  We present a new approach to foreground removal for Cosmic Microwave
  Background (CMB) maps. Rather than relying on prior knowledge about
  the foreground components, we first extract the necessary
  information about them directly from the microwave sky maps by
  taking differences of temperature maps at different
  frequencies. These difference maps, which we refer to as internal
  templates, consist only of linear combinations of galactic
  foregrounds and noise, with no CMB component. We obtain the
  foreground cleaned maps by fitting these internal templates to, and
  subsequently subtracting the appropriately scaled contributions of
  them from, the CMB dominated channels. The fitting operation is
  performed in wavelet space, making the analysis feasible at high
  resolution with only a minor loss of precision. Applying this
  procedure to the \emph{WMAP} data, we obtain a power spectrum that
  matches the spectrum obtained by the \emph{WMAP} team at the signal
  dominated scales. The fact that we obtain basically identical
  results without using any external templates has considerable
  relevance for future observations of the CMB polarization, where
  very little is known about the galactic foregrounds. Finally, we
  have revisited previous claims about a north-south power asymmetry
  on large angular scales, and confirm that these remain unchanged
  with this completely different approach to foreground
  separation. This also holds when fitting the foreground contribution
  independently to the northern and southern hemisphere indicating
  that the asymmetry is unlikely to have its origin in different
  foreground properties of the hemispheres. This conclusion is further
  strengthened by the lack of any observed frequency dependence.

\end{abstract}


\section{Introduction}
\label{sect:intro}

The \emph{Wilkinson Microwave Anisotropy Probe} (\emph{WMAP})
satellite \citep{bennettwmap} and other recent ground-based and
balloon-borne experiments have provided high sensitivity observations
of the Cosmic Microwave Background (CMB). High-precision estimates of
the cosmological parameters have been derived from these data, and
these will improve yet further with data from Planck and near-future
ground based experiments. Later still, high sensitivity polarization
data of the CMB will be available and the physics of the early
universe can be studied with even higher fidelity. However, in order
to obtain reliable estimates of the cosmological parameters, control
of systematic effects and the different sources of foreground
contamination is of the utmost importance.  In this paper we address
the latter problem.

There are three well-understood sources of diffuse galactic
contamination.  Synchrotron emission from the Galaxy originates in
relativistic cosmic ray (CR) electrons spiraling in the Galactic
magnetic field.  Free-free emission is the bremsstrahlung radiation
resulting from the Coulomb interaction between free electrons and ions
in the Galaxy.  Thermal dust emission arises from grains large enough
to be in thermal equilibrium with the interstellar radiation field.
Evidence of an additional component initially arose from the
cross-correlation of the \emph{COBE}-DMR data with the DIRBE map of
thermal dust emission at 140$\mu$m \citep{kogut:1996} that revealed
anomalous emission with a spectrum rising as the frequency decreased
from 53 to 31.5 GHz. Both \citet{banday:2003} and
\citet{bennettgalaxy} suggested that this component was characterized
by a power-law frequency spectrum with index -2.5 over the range
$\sim$20 - 60 GHz.  However, it remains unclear what the physical
origin of such emission is -- \citet{bennettgalaxy} have proposed that
it arises from hard synchrotron emission in star-forming regions close
to the Galactic plane, whilst \citet{draine:1998} suggest rotational
emission from very small grains or `spinning dust'.  Recently,
\citet{watson:2005} have shown that observations of the Perseus
molecular cloud between 11 and 17 GHz and augmented with the
\emph{WMAP} data can be adequately fitted by a spinning dust model.
Nevertheless, we will refer to the putative component as anomalous
dust.

Fortunately, the foreground components have frequency spectra very
different from the CMB, although the exact shapes of these spectra are
not known. Most data on galactic emission have been taken at
frequencies others than those used for CMB observations and it is not
clear whether using these as templates for CMB foreground subtraction
is a valid approach. Certainly spatial variations in the frequency
dependence of the foregrounds will cause the templates to increasingly
diverge from the true structure at microwave wavelengths.
Nevertheless, given the lack of other reliable approaches, a template
fitting and subtraction procedure was applied on the \emph{WMAP} data
\citep{bennettwmap} using external templates
\citep{bennettgalaxy,hinshaw}. The Spectral Matching Independent
Component Analysis (SMICA) \citep{smica} method has been applied to
the \emph{WMAP} maps after they had already been cleaned using the
standard external templates and evidence for a residual galactic
component was found \citep{smicawmap}. The \emph{WMAP} data were also
corrected using the Internal Linear Combination (ILC) methods
\citep{bennettgalaxy, ilc2, ilc3} but unfortunately these methods are
biased \citep{ilc3}. Other methods that have been developed for
foreground subtraction
\citep{mem1,mem2,mem3,mem4,ica1,ica2,ica3,brandt,fgfit}
have yet to be applied to the \emph{WMAP} data.

The importance of being able to make reliable foreground corrections
on large angular scales with limited knowledge about the foreground
components becomes even more apparent when considering future CMB
polarization observations. A very important test of inflation depends
on high precision measurements of the predicted B-mode polarization of
the CMB on large angular scales. However, very little is known about
the polarization of the galactic emission. Future high sensitivity
measurements of the CMB polarization are likely to be highly dependent
on foreground subtraction methods that do not make strong assumptions
about the nature of the galactic emission.

Here we propose a new foreground subtraction method for which no
knowledge about the morphology or frequency spectra of galactic
components are required.  The method does not require constant (in
frequency) spectral indexes for the galactic emission components and
will therefore not be biased by uncertainties in the changes of the
spectral indexes over large frequency ranges. It will however require
the spectral indexes to be constant in space over a given patch on the
sphere. In the implementation presented here, we assume the spectral
indexes to be constant over the full sphere or over hemispheres but
the extension to smaller patches will be discussed. As the method
increases the noise level in the data, it is at present mainly useful
for studies of the larger scales where noise is not dominant.  This is
particularly relevant given that the largest angular scales in the
\emph{WMAP} data have indicated the presence of several anomalies, in
particular a north-south asymmetry in the power spectrum
\citep{asymm1,asymm2}. This result will be re-examined here with our
new foreground corrected map.

The method presented here takes advantage of the fact that the
observed data already provide information about the foregrounds. In
particular, by computing differences of maps observed at different
frequencies, (noisy) linear combinations of the foregrounds components
are obtained. Fitting and subtracting such templates that are simple
linear combinations of the foregrounds is mathematically equivalent to
fitting and subtracting templates of the physical foreground
components. The advantage of such linear combinations is that they can
be obtained directly from the microwave observations themselves and
therefore do not rely on other experiments. As a consequence, the
foreground morphologies are likely to be well traced even in the
presence of modest departures from a single spectral index in a given
region.

In \S \ref{sect:idea}, we present basis of the method and demonstrate
its application in detail. Then, in \S \ref{sect:test} and \S
\ref{sect:wmap} the results of this procedure as applied to both
simulated inputs and the \emph{WMAP} sky maps are presented. Finally,
in \S \ref{sect:concl}, we discuss how the method can be improved in
the future.

\section{Methodological Basis: Fitting and Subtracting Internal Templates}
\label{sect:idea}

A pixelized temperature map of microwave observations can be written (in
thermodynamic temperatures) as
\begin{equation}
\label{eq:ti}
T_i^\nu=T_i^\mathrm{CMB}+n_i^\nu+\sum_{t=1}^{N_t}c_t^\nu s_i^t.
\end{equation}
Here $T_i^\nu$ is the observed temperature in pixel $i$ for frequency
channel $\nu$, $T_i^\mathrm{CMB}$ is the frequency independent CMB
component, $n_i^\nu$ is the instrumental noise and finally $s_i^t$ is
the contribution from galactic foreground component $t$ for pixel $i$.
We assume a total of $N_t$ different foreground components. The
coefficients $c_t^\nu$ give the amplitude of the given foreground
component $t$ in channel $\nu$. In this paper we assume spatially
constant foreground frequency spectra, as was also assumed by the
\emph{WMAP} team in generating the publicly available foreground cleaned maps.
However, as discussed in the conclusions, the method 
can be extended to take into account variations of spectral
indexes across the sky, but implementation of this is deferred to a
future publication. 

\subsection{Internal templates}

The \emph{WMAP} team used external templates obtained from
observations at other frequencies as tracers of the foreground
components $s_i^t$.  The corresponding coefficients $c_t^\nu$ were
then obtained by a fitting procedure (the details of which are not
clear) and the inferred foreground contributions subsequently
subtracted from the sky maps. Here, we will not use external
templates, but instead take the difference between two frequency
channels. In such difference maps, which we call {\it internal
templates}, the CMB cancels out and the remainder is a linear
combination of foreground components plus instrumental noise.

We can write the internal templates $D^{\nu\nu'}_i$, as
\begin{equation}
D^{\nu\nu'}_i\equiv T_i^\nu-T_i^{\nu'}=\sum_{t=1}^{N_t}(c_t^\nu-c_t^{\nu'})s_i^t+\delta n_i^{\nu\nu'},
\end{equation}
where $\delta n_i^{\nu\nu'}=n_i^\nu-n_i^{\nu'}$ is the noise of the
internal template. Note that we have assumed the maps $T_i^\nu$ at all
frequencies to have been smoothed to the same resolution.

The natural physical `basis' for the foreground components may be
identified with dust, synchrotron and free-free emission.  The
internal templates are transformations into a new basis consisting of
linear combinations of the physical components. Fitting and
subtracting foreground components in this new basis is clearly
mathematically equivalent to using the 'natural basis', provided that
the number of internal templates is equal to the number of components.
Therefore, using the internal templates $D^{\nu\nu'}_i-\delta
n_i^{\nu\nu'}$ as our new basis, equation (\ref{eq:ti}) can be written
as
\begin{equation}
\label{eq:newti}
T_i^\nu=T_i^\mathrm{CMB}+n_i^\nu+\sum_{t=1}^{N_t}\tilde c_t^\nu(D_i^t-\delta n^t),
\end{equation}
where the sum is performed over internal templates $t=\nu\nu'$ for
$N_t$ different combinations of channels.

It is important to note that because the internal templates are noisy,
subtracting them from the data will increase the overall noise level
of the final map and significantly increase the error bars for power
spectrum estimates at the highest multipoles.

\subsection{Fitting procedure}

When using external templates $\hat s_i^t$ for the different
foreground components, the fitting procedure is performed by
minimizing the $\chi^2$,
\begin{equation}
\label{eq:chi2}
\chi^2=\sum_{ij}\sum_{\nu}(T_i^\nu-\sum_tc_t^\nu\hat s_i^t)C^{-1}_{i,j}(T_j^{\nu}-\sum_tc_t^{\nu}\hat s_j^t)
\end{equation}
for the coefficients $c_t^\nu$.  Here the sum is performed over pixels
$i$ and $j$ as well as channels $\nu$ including only the
channels we wish to clean. The correlation matrix is defined by
$C_{i,j}=\VEV{T_i  T_j}-\VEV{T_i}\VEV{T_j}$ which can be obtained by
Monte-Carlo simulations assuming a reasonable CMB spectrum. 
In this paper we use a similar fitting procedure,
replacing $\hat s_i^t\rightarrow D_i^t$. In principle we should have
used $D_i^t-\delta n_i^t$ (see equation \ref{eq:newti}), but since the
noise is known only in a statistical sense we are forced to use
$D_i^t$ and correct for the resulting noise bias. This bias correction
procedure will be described in detail in Appendix \ref{sect:app}.

In practise, the evaluation of equation (\ref{eq:chi2}) is not
feasible for high resolutions. The covariance matrix is too large to
be stored in a computer and too time consuming to invert. For this
reason we adopt a wavelet space implementation of the above
procedure. The advantage is that we can neglect pixel-pixel
correlations, taking only into account the scale-scale correlations in
the correlation matrix without significant loss of precision.

We adopt the wavelet fitting procedure of, e.g.,
\citet{vielva} and \citet{hansensz} and define the cross-correlation coefficients
\begin{equation}
\label{eq:xdef}
X^{\nu t}_S\equiv\sum_iw_{iS}^\nu w_{iS}^t,\ \ \ X^{tt'}\equiv\sum_iw_{iS}^tw_{iS}^{t'}.
\end{equation}
Here $w_{iS}^\nu$ is the wavelet transform of $T_i^\nu$ for pixel $i$,
channel $\nu$ and scale $S$, and $w_{iS}^t$ is the wavelet transform
of the internal template $D_i^t$. We use Spherical Mexican Hat
wavelets \citep{wavelets} which are suitable for both template fitting
\citep{vielva,hansensz}, point source detection \citep{vielvapoint}
and tests of non-Gaussianity \citep{vielvaspot,wang,cabella}.

In wavelet space, the $\chi^2$ to minimize has the form
\begin{equation}
\chi^2=\sum_{\nu T}\sum_{SS'}(X_S^{\nu T}-\sum_t\tilde c_t^\nu X_S^{tT})C_{SS'}^{-1}(X_{S'}^{\nu T}-\sum_t\tilde c_t^\nu X_{S'}^{tT}),
\end{equation}
where the summation runs over all target channels $\nu$, all internal templates
(foreground components) $T$ and all wavelet scales $S$ and $S'$. The
scale-scale correlation matrix $C_{SS'}$ is obtained by Monte-Carlo
simulations. Minimizing the $\chi^2$, the best fit $\tilde c_t^\nu$  are given by a set of equations
\begin{equation}
\sum_t \tilde c_t^\nu M_{tf}=B_f^\nu
\end{equation}
for each template $f$ and each frequency $\nu$. Here the matrix $M_{tf}$ is given by
\begin{equation}
 M_{tf}=\sum_T\sum_{SS'}X_S^{fT}C_{SS'}^{-1}X_{S'}^{tT}.
\end{equation}
and the vector $B_f^\nu$ by
\begin{equation}
B_f^\nu=\sum_T\sum_{SS'}X_S^{fT}C_{SS'}^{-1}X_{S'}^{\nu T}
\end{equation}

An estimate of the error bars on the coefficients $\tilde c_t^\nu$ can
be obtained performing Monte-Carlo simulations of CMB, noise and
`foregrounds'. In these simulations, the internal templates obtained
from the data are used as `foregrounds' with the best estimate
coefficients obtained from the data. An additional correction to the
$X_S$ is necessary in this case as described in Appendix
\ref{sect:app}. The error bars obtained in this manner
has been shown to agree well with the real error bars.

The wavelet scales to use in the analysis are chosen such that the
scale-scale correlation matrix is well conditioned. In order to
determine these scales, we adopt the following procedure:
\begin{itemize}
  \item Use MC simulations to obtain $C_{SS'}$ for a large set of scales.
  \item Define a limit $\alpha$, say, 0.95. Start by the smallest
    allowed scale $S$ and find the next scale by identifying at which
    scale the normalized covariance matrix
    $C_{SS'}/\sqrt{C_{SS}C_{S'S'}}$ has fallen to $\alpha$.
  \item Repeat the above procedure until the largest scale has been
    reached. Check whether the final correlation matrix is well
    conditioned. If not, decrease $\alpha$ and repeat.
\end{itemize}

As in pixel space, a correction procedure for the noise bias has to be
implemented. In Appendix \ref{sect:app} we will describe this
procedure in detail as well as a procedure for estimating the level of
the remaining bias.

\section{Testing the Method on Simulations}
\label{sect:test}

The foreground subtraction procedure described above, which we will
call Wavelet based hIgh resolution Fitting of Internal Templates
(WI-FIT), has been extensively tested on simulated maps. We generated
a set of 500 Monte-Carlo simulations of CMB and noise, using the
best-fit \emph{WMAP} power-law power spectrum and noise properties
corresponding to the five \emph{WMAP} channels\footnote{These may be
obtained from the Lambda website {\em http://lambda.gsfc.nasa.gov/}.}.
To these simulations, we added contributions from known galactic
foregrounds based on specific templates. In particular, for thermal
dust we use the template provided by \citet{schlegel} extrapolated in frequency by \citet{fink:1999}, for synchrotron we adopt the
408 MHz sky survey of \citet{haslam:1982}, and for the free-free
contribution we assume that the emission can be traced by a template
of H$_\alpha$ emission \citep{fink:2004}. The templates are then
scaled to the \emph{WMAP} frequencies using the weights given in Table
4 of \citet{bennettgalaxy}.  These dust weights effectively include an
anomalous dust contribution assuming that the putative emission can
also be well traced by the thermal dust template.  

In order to validate the wavelet basis of our high resolution
analysis, we first make an explicit comparison of the method to a full
pixel space foreground subtraction procedure at lower resolution where
it remains feasible.  The maps were therefore smoothed to a common
resolution of $5.5^\circ$ and degraded to
HEALPix\footnote{http://healpix.jpl.nasa.gov/} resolution
$N_\mathrm{SIDE}=32$.  Unless otherwise stated, the \emph{WMAP} Kp2
sky cut was applied in all the analyses presented in this paper.
Using the procedure with internal templates as described above, then
after including the bias correction we obtained unbiased estimates of
the foreground coefficients $\tilde c_t^\nu$ in pixel space. Repeating
the same procedure in wavelet space yields approximately $2-3\%$
larger error bars showing that the loss in precision using this
approach instead of the full pixel space procedure is
small. Therefore, since analysis at higher resolution is in any case
unfeasible in pixel space, we adopt the wavelet approach in what
follows.
 
\begin{figure*}
\begin{center}
\includegraphics[width=0.8\linewidth]{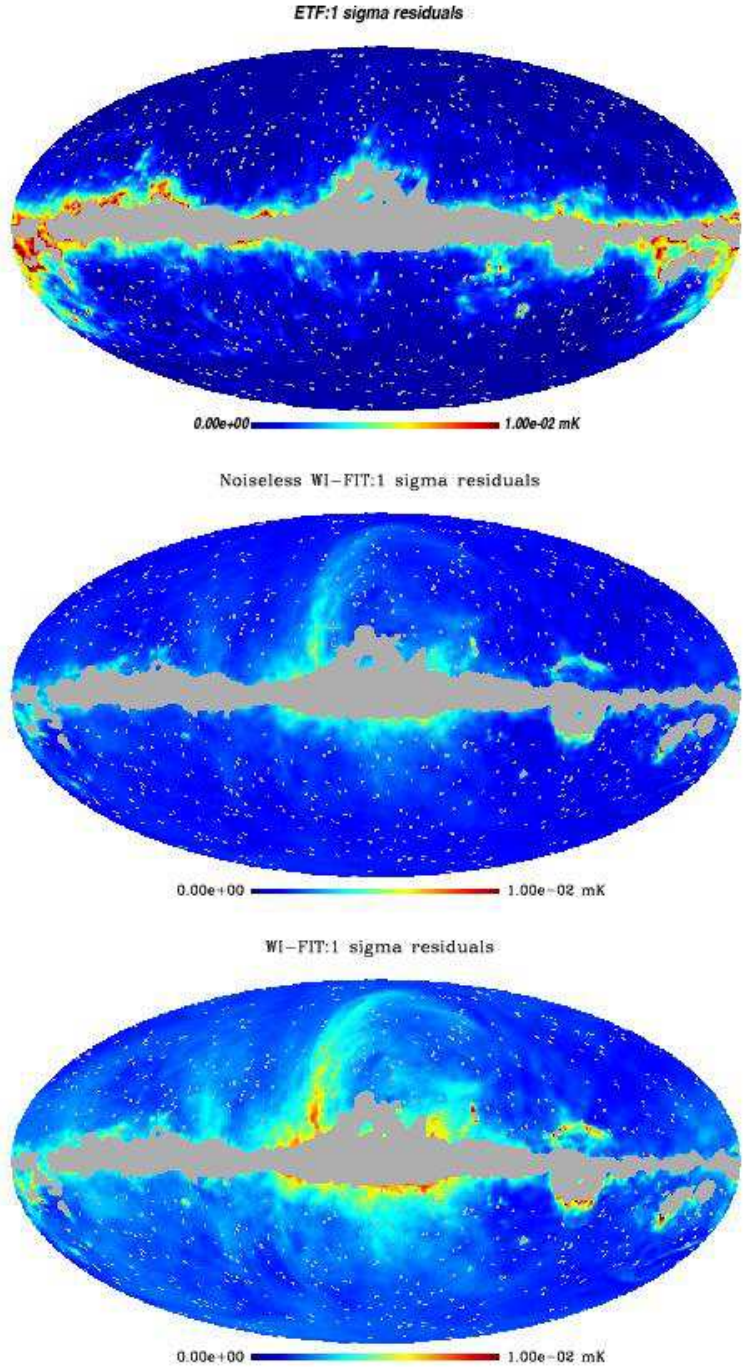}
\caption{ Results from 500 Monte-Carlo simulations: The upper plot
shows the $1\sigma$ residuals $\sqrt{\VEV{\delta_i^2}}$ on the V-band
from the wavelet fitting procedure using external templates. The
middle and lower plot show the same result when using internal
templates without and with noise respectively. Note that in these
simulations the external templates are equal to the foregrounds added
to the simulated maps. In reality, templates taken at frequencies
different from the ones dominated by the CMB will differ from the
foregrounds contaminating the CMB. This additional error is not
reflected in the upper plot. Note further that the last case generally
has higher noise variance than the two other cases, but the noise
variance has not been included in the plots in order to make the
foreground residuals clearer. \label{fig:stdev}}
\end{center}
\end{figure*}

We then attempted to calibrate the efficiency of internal versus
external template fitting.  The WI-FIT procedure was applied to maps
of 1 degree FWHM resolution at $N_\mathrm{SIDE}=256$ and compared to
wavelet-based fitting of external templates.  The results derived
using 13 wavelet scales is shown in Fig. \ref{fig:stdev}. The two
upper maps show the $1\sigma$ residuals for the external and internal
template subtraction method when the internal templates were
noise-free. These maps $\sigma_i$ were obtained by
\[
\sigma_i\equiv\sqrt{\VEV{\delta_i^2}}\equiv\sqrt{\frac{1}{N_s}\sum_s(M^s_i-T_i^s)^2},
\]
where the sum is performed over all simulations $s$, in total
$N_s=500$, $M^s_i$ is the cleaned map for pixel $i$ and realization
$s$ and $T_i^s$ is the input CMB realization.

Unfortunately, the exercise is not particularly revealing, and mostly
emphasises that external templates are likely to outperform internal
templates when we know that they accurately describe the foregrounds
present in the data.  In the example shown here, the internal
templates do better close to the galactic plane, but worse
outside. Additionally, since the internal templates will be noisy, the
$1\sigma$ errors will be further inflated as shown in the lower map.
Furthermore, the exact level of residuals will in general depend on
the choice of internal templates (the specific difference combinations
adopted in the analysis) as well as the nature of the foregrounds.  In
fact, the comparison somewhat violates the philosophy underpinning the
WI-FIT method.  The external templates are measured at frequencies
different from those used for studies of the CMB and are therefore not
likely to accurately trace the foregrounds in the CMB dominated
channels (although they do by construction in this test).  The
internal templates however, are linear combinations of the foregrounds
at the channels used for CMB studies.  The comparison is therefore not
completely fair and the results shown in the plot do not adequately
reflect a true comparison.  We nevertheless present the results for
completeness.

\section{Application to the \emph{WMAP} Data}
\label{sect:wmap}

\subsection {The maps}

We have applied the WI-FIT method to the first-year \emph{WMAP} data
smoothed to 1 degree FWHM at $N_\mathrm{SIDE}=256$ and using the Kp2
pixel mask. For each of the \emph{WMAP} sky maps covering five
frequency bands -- K, Ka, Q, V and W (when multiple maps are available
at the same frequency we take simple averages of the maps after
convolution to the 1 degree FWHM beam) --
three internal templates were constructed using the four remaining
bands. For example, for the Q-band the internal templates K--Ka, Ka--V
and V--W were generated, fitted to the Q-band sky map, and subtracted
according to the above prescription. 

\begin{figure*}
\begin{center}
\includegraphics[width=\linewidth]{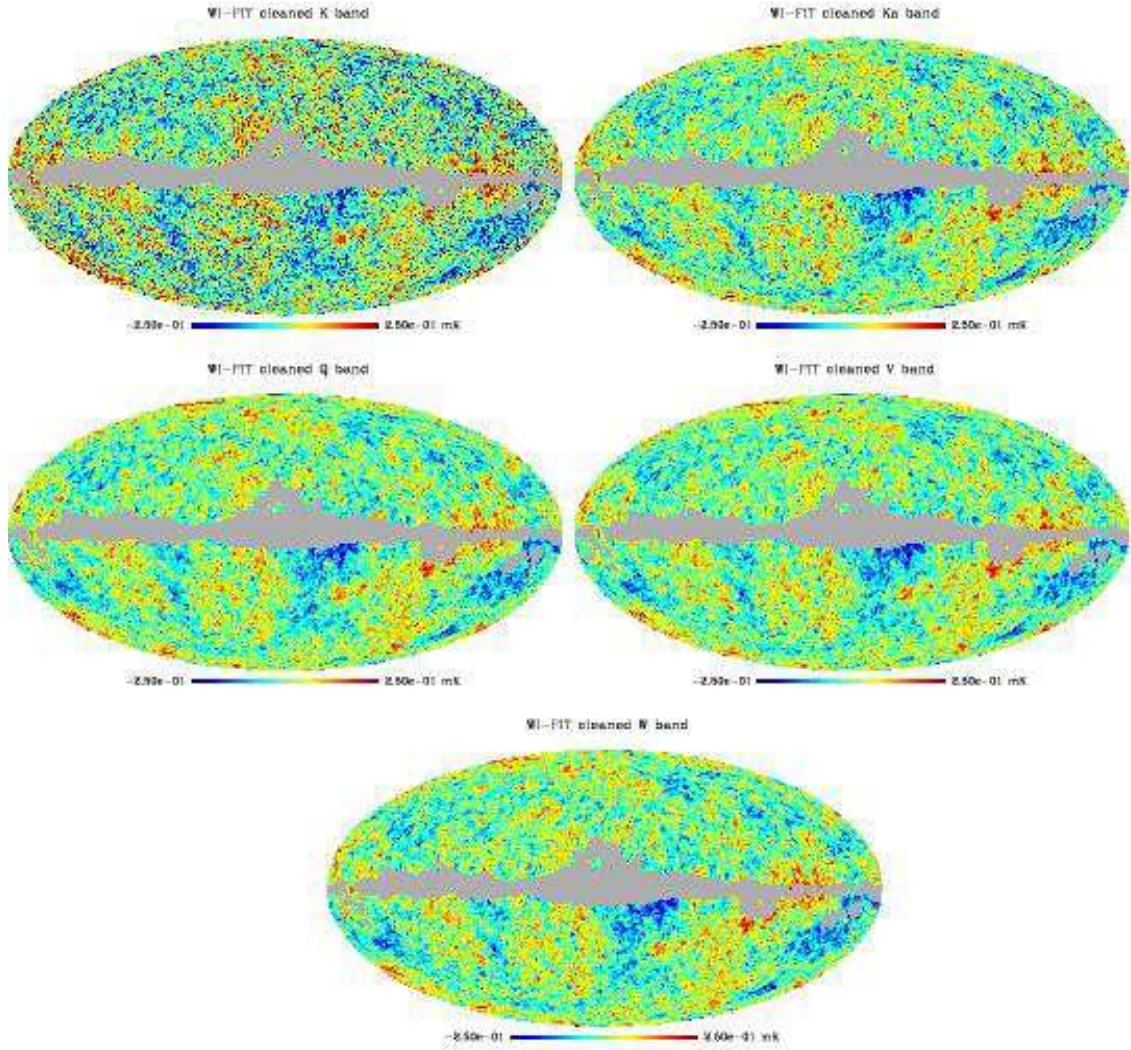}
\caption{ WI-FIT cleaned \emph{WMAP} maps: The 5 \emph{WMAP} channels
smoothed to a common resolution of 1 degree FWHM and cleaned with the
WI-FIT procedure applied outside the Kp2 galactic
cut. \label{fig:allmaps}}
\end{center}
\end{figure*}

In Fig. \ref{fig:allmaps}, we show the cleaned K, Ka, Q, V and W
maps. The template coefficients for the K band are so large that the
noise level of the resultant cleaned map is too high to be of
practical value in any further analysis and will not be considered in
what follows. The maps shown can be obtained by combining the
different \emph{WMAP} bands according to the weights given in table
\ref{tab:weights}.  In the upper plot of Fig. \ref{fig:coaddedmaps} we
also show a noise-weighted combination of the Q, V and W bands. We do
not present error bars here as these error bars only are valid when
the model (that the spectral indexes of the foregrounds are the same
in all directions) is valid. We know that the assumption of spatially
constant spectral indexes is wrong and the error bars would hence be
misleading.

\begin{figure*}
\begin{center}
\includegraphics[width=\linewidth]{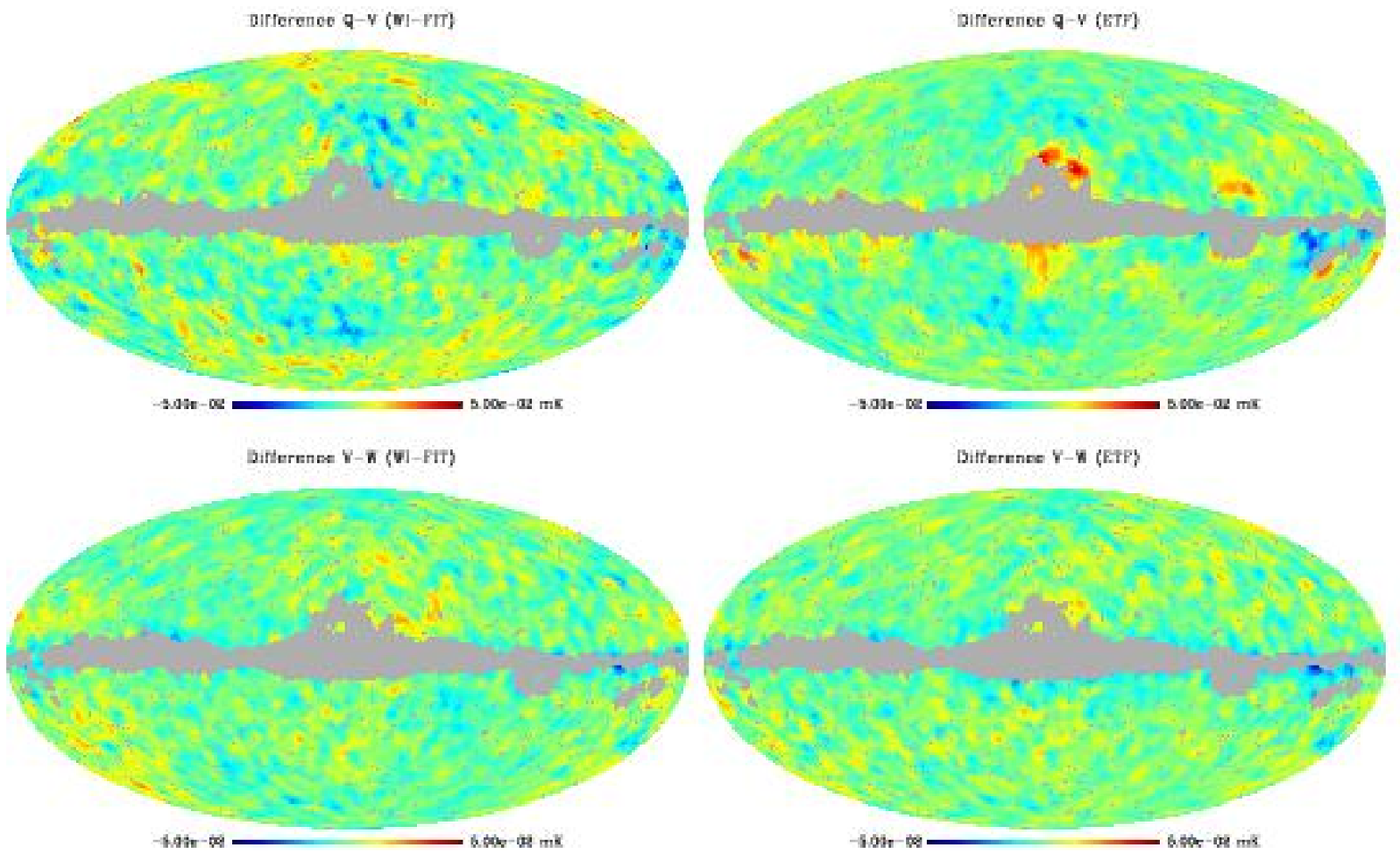}
\caption{ The plots in the left column show the median filtered (3 degree
radius top-hat) differences Q-V and V-W for the WI-FIT cleaned \emph{WMAP}
maps where only that part of the sky outside the Kp2 mask was used in the
fitting procedure. The plots in the right column show the same differences for
the maps that have been foreground corrected by the \emph{WMAP}
collaboration using external templates. \label{fig:diffmaps}}
\end{center}
\end{figure*}

In Fig. \ref{fig:allmaps} there are no obvious differences between the
Ka, Q, V and W bands. In order to check for possible foreground
residuals, we have taken the differences between these sky maps.
Since these foreground residuals are generally smaller than the
instrumental noise level, the difference maps resemble pure noise
without further processing.  For visualisation purposes, we have
applied a median filter with a 3 degree radius to suppress the noise
on pixel scales.  Fig. \ref{fig:diffmaps} shows the the difference
maps for Q--V and V--W. For a perfect foreground subtraction, the
difference maps should show little coherent structure beyond that
expected for median-filtered pure noise.  It is apparent that some
small residuals remain outside the Kp2 cut.  For comparison, the
corresponding differences of the \emph{WMAP} maps cleaned by External
Template Fitting (ETF) provided by the \emph{WMAP} collaboration are
also shown. The ETF maps seem to have stronger residuals than the
WI-FIT maps close to the galactic cut. The WI-FIT maps show stronger
fluctuations over the whole sky, but this can partly be explained by
the higher noise level in these maps.

\begin{figure*}
\begin{center}
\includegraphics[width=0.8\linewidth]{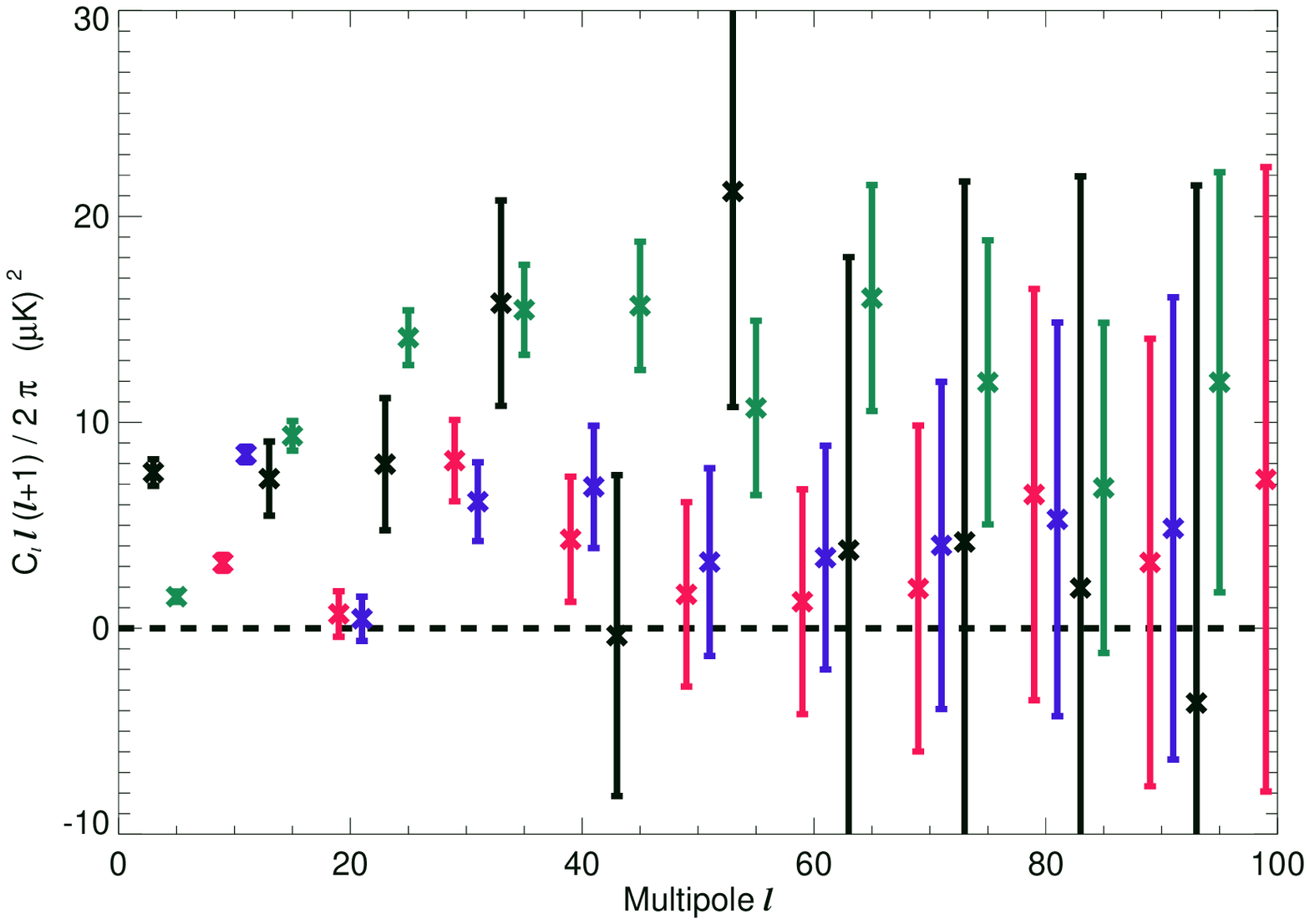}
\caption{ The binned power spectra of the difference maps Q--V and V--W for the ETF and WI-FIT maps (The Kp2 galactic cut was used in the analysis). Black points: WI-FIT Q--V, green points: ETF Q--V, red points: WI-FIT V--W, blue points: ETF V--W. The error bars are obtained from pure noise simulations. \label{fig:diffspectra}}
\end{center}
\end{figure*}

In Figure (\ref{fig:diffspectra}) we show the noise-corrected spectra of these four difference maps. For the Q--V difference, WI-FIT shows stronger residuals in the lowest multipole bin, but ETF has residuals well above $\ell=40$ where the WI-FIT difference spectrum is consistent with pure noise. For the V--W map however, ETF shows stronger residuals in the smallest multipole bin whereas for all other bins the two maps are consistent.

Note further that the Q--V ETF difference map shows strong
similarities to the residual component obtained by the SMICA
method. \citet{smicawmap} found evidence for a residual galactic
component (their Figure 5) mainly present in the Q band but not
detected in the V band\footnote {This feature is actually faintly
visible in Figure 11 of \citet{bennettgalaxy}.}. It is therefore not
unexpected that this component shows up when taking the difference
Q--V.

In the V--W difference maps one can in both cases see a blue region
following the border of the galactic cut (this is more prominent in
the ETF maps). As the W band is mainly contaminated by thermal dust,
this can be interpreted as a sign of residual thermal dust
contamination. That there is a residual thermal dust contamination in
the W band is supported by table \ref{tab:weights} where we have
assumed a spectral index for each foreground component
($\beta_{\mathrm{s}} = -2.7$ for synchrotron, $\beta_{\mathrm{ff}} =
-2.15$ for free-free, $\beta_{\mathrm{d}} = 2.2$ for thermal dust and
$\beta_\mathrm{sd}=-2.5$ for anomalous dust) and estimated the
proportions of the residual foreground components (see \citet{ilc3}
for details). These numbers indicate the fraction of residual
foreground contribution for each type relative to the uncorrected
level at an adopted reference frequency (22.8 GHz for synchrotron and
anomalous dust, 33.0 GHz for free-free and 93.5 GHz for thermal dust).

\begin{figure*}
\begin{center}
\includegraphics[width=\linewidth]{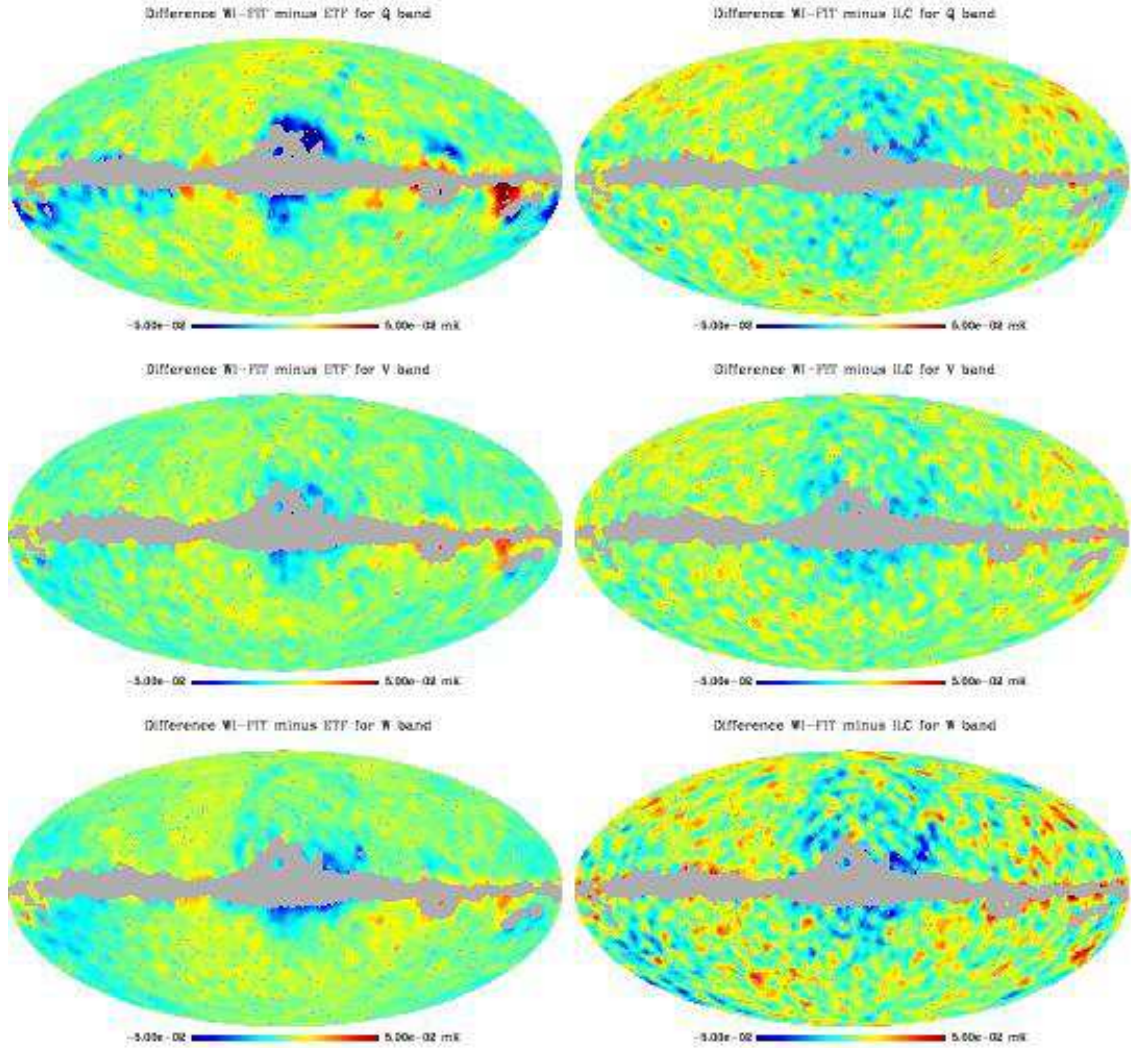}
\caption{ The plot shows the median filtered (3 degree radius top-hat)
differences between the WI-FIT and ETF/ILC maps obtained by the
\emph{WMAP} collaboration for the three frequencies Q, V and
W. \label{fig:diffwmap}}
\end{center}
\end{figure*}

In Fig. \ref{fig:diffwmap} we show the filtered difference maps
between the WI-FIT and ETF maps and the ILC map provided by the
\emph{WMAP} collaboration. The Figures suggest the presence of
residuals at the $10\%$ level compared to the CMB over the full sky in
either the WI-FIT maps, ETF maps or both. Note that the
\emph{WMAP}-ETF difference for the Q band shows similarities to the
residual component detected with SMICA. The fact that the hot and cold
areas are interchanged indicates that the residual is not present in
the WI-FIT cleaned maps. A ring-like structure on large angular scales
is also in all channels. It is not clear whether this residual
structure can be attributed to the ETF or the WI-FIT cleaned maps. As the residual is positive in WI-FIT minus ETF, it can either be a strucure which is corrected for in ETF but not in WI-FIT, or an overcorrection in ETF. Looking at the power spectra of these difference maps (Fig. \ref{fig:cl_wmap}), the residual does not significantly affect the estimated CMB power spectra.

\subsection{The power spectrum}

\begin{figure*}
\begin{center}
\includegraphics[width=0.7\linewidth]{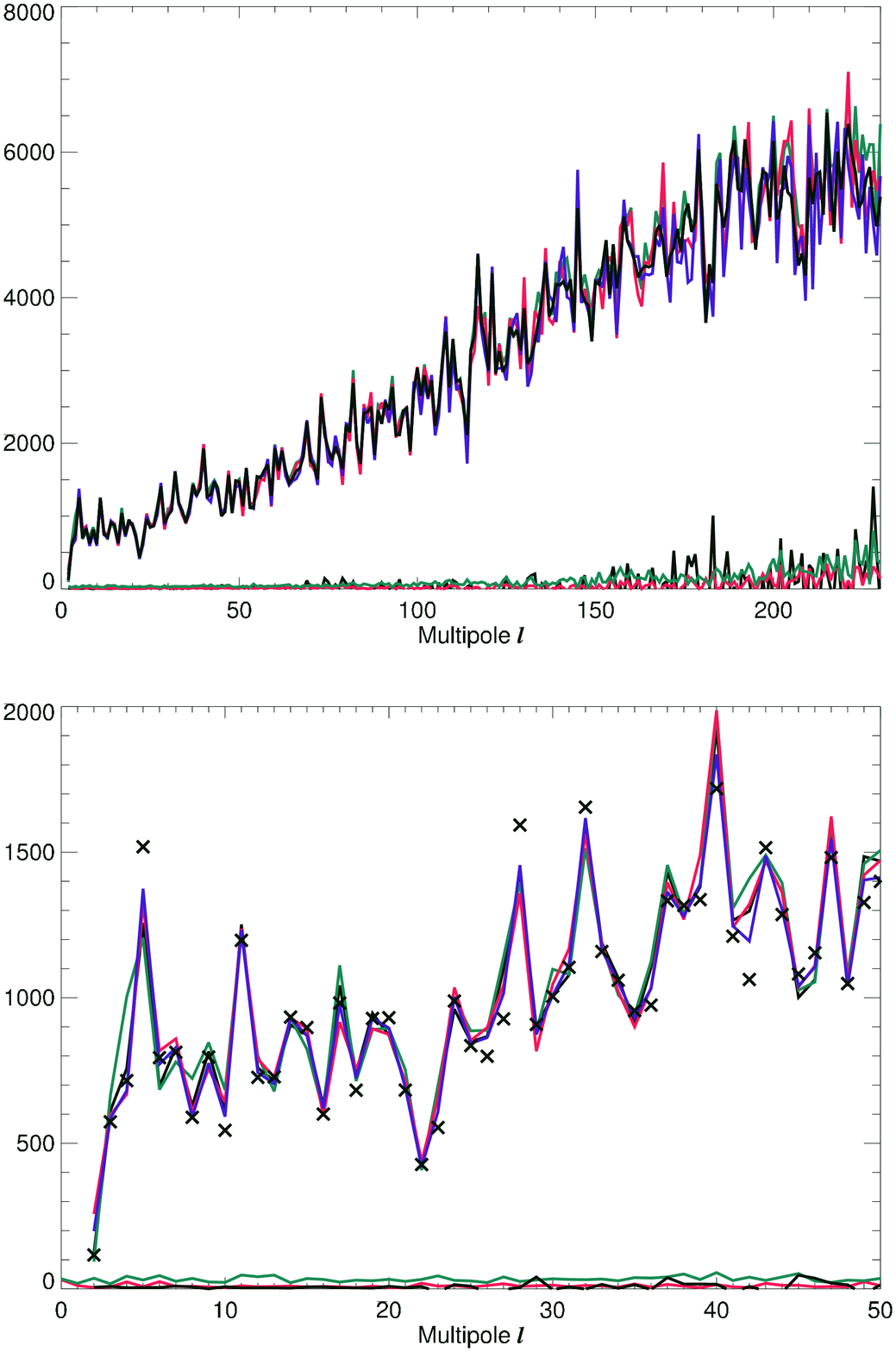}
\caption{ The power spectra of the WI-FIT cleaned \emph{WMAP} maps for
different bands compared to the power spectrum obtained by the
\emph{WMAP} team (solid black line). Green line shows the result for
the Q band, blue line for the V band, red line for the W band and
black crosses for the Ka band. The lower green and red lines show the
power spectra of the difference WI-FIT minus ETF for the Q and W band
respectively. The power spectrum of the difference map Q-W for the
WI-FIT maps is shown as a black line.\label{fig:cl_wmap}}
\end{center}
\end{figure*}

\begin{figure*}
\begin{center}
\includegraphics[width=0.8\linewidth]{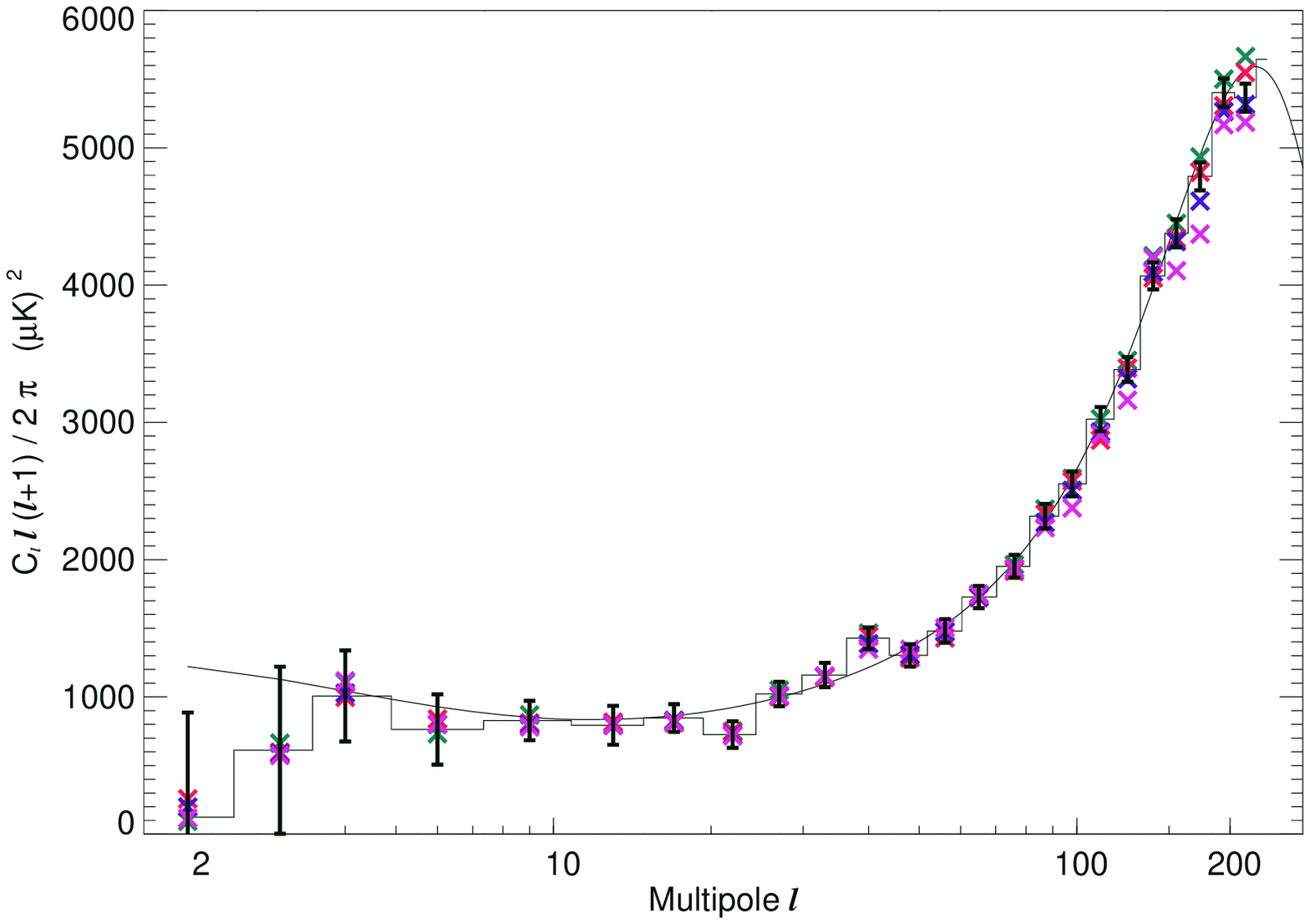}
\caption{ The binned power spectra of the WI-FIT cleaned \emph{WMAP} maps for
different bands compared to the power spectrum obtained by the
\emph{WMAP} team (histogram and error bars). Green crosses show the result for
the Q band, blue crosses for the V band, red crosses for the W band and
pink crosses for the Ka band.\label{fig:cl_binned}}
\end{center}
\end{figure*}

In this section, we consider the implications of these residuals for power
spectrum estimation.  The MASTER algorithm \citep{master} was applied 
to the WI-FIT foreground cleaned maps outside the Kp2 sky cut.
Before obtaining the spectra, the best fit mono- and dipole were
subtracted from all maps. Fig. \ref{fig:cl_wmap} presents the
results (binned version in Fig. \ref{fig:cl_binned}). For comparison, the power spectrum estimated by
the \emph{WMAP} collaboration \citep{hinshaw} from the ETF cleaned maps using
the cross power spectrum is shown by a solid black line. 
The WI-FIT spectra are shown as green (Q),
blue (V) and red (W) lines and the Ka spectrum is delineated by crosses.

There is good agreement between the V and W band spectra estimated
with the WI-FIT blind approach and the ETF cleaning method based on external
observations of the galaxy at very different frequencies.  Small
differences in the Q band spectrum probably indicate residual
foreground contamination, more obviously present in the results for
the highly foreground contaminated Ka channel.  Nevertheless, both
agree reasonably well with the V and W band spectra.
Finally, we also show noise-corrected spectra of the 
WI-FIT and ETF difference maps. 
The foreground residuals outside the Kp2 cut which show up when 
taking the difference between the WI-FIT cleaned Q, V and W maps 
(Fig. \ref{fig:diffmaps}) as well as the difference between the ETF and WI-FIT
maps (Fig. \ref{fig:diffwmap}) are clearly unlikely to cause significant
differences in the CMB power spectrum.

\begin{deluxetable}{cccccccccc}
\tabletypesize{\small}
\tablecaption{\emph{WMAP} band weights for the WI-FIT cleaned maps and
approximate foreground residuals assuming the following spectral
indices for the foreground components:
$\beta_\mathrm{synchrotron}=-2.70$, $\beta_\mathrm{free-free}=-2.15$,
$\beta_\mathrm{thermal dust}=2.20$ and $\beta_\mathrm{anomalous
dust}=-2.5$. The residuals are given relative to the expected synchrotron and
anomalous dust level at 22.8 GHz, free-free level at 33 GHz and
thermal dust level at 93.5 GHz. \label{tab:weights}}
\tablewidth{0pt}
\tablehead{
\colhead{\emph{WMAP} band/region} & \colhead{K} & \colhead{Ka} & \colhead{Q} & \colhead{V} & \colhead{W} & \colhead{synch.} & \colhead{ff} & \colhead{th. dust} & \colhead{an. dust} }
\startdata

K  &  1.00 & -3.75 &  0.83 &  3.81 & -0.90 & 0.04 & -0.10 & 0.42 & 0.00\\
Ka  & -0.17 & 1.00 & -1.91 &  2.10 & -0.02 & -0.06 & -0.05 & 0.52 & -0.05\\
Q  & -0.02 & -0.75 &  1.00 &  0.34 &  0.44 & -0.06 & 0.05 & 0.55 & -0.05 \\
Q (north) &  -0.05 & -0.68 &  1.00 &  0.50 &  0.22 & -0.06 & -0.02 & 0.44 & -0.04 \\
Q (south) &   0.46 & -2.27 &  1.00 &  2.37 & -0.57 & -0.02 & -0.06 & 0.33 & -0.02 \\
V  & -0.23 & 0.66 & -0.77 &  1.00 &  0.33 & -0.07 & -0.06 & 0.55 & -0.06 \\
V (north) &  -0.33 &  1.35 & -1.55 &  1.00 &  0.53 & -0.08 & -0.06 & 0.66 & -0.06 \\
V (south) &   0.21 & -1.64 &  1.35 &  1.00 &  0.07 & -0.04 & -0.06 & 0.46 & -0.04\\
W  & -0.34 & 0.76 & -0.18 & -0.24 &  1.00 & -0.10 & -0.09 & 0.75 & -0.08\\
W (north) &  -0.62 &  3.10 & -3.49 &  1.01 &  1.00 & -0.12 & -0.14 & 0.90 & -0.10 \\
W (south) &  -0.03 & -1.45 &  2.61 & -1.13 &  1.00 & -0.08 & -0.07 & 0.66 & -0.07 \\
co-added  &  -0.17 &  0.06 &  0.17 &  0.38 &  0.56 & -0.08 & -0.07 & 0.61 & -0.06 \\
co-added (north)   & -0.04 & -0.64 &  0.85 &  0.39 &  0.43 & -0.07 & -0.06 & 0.55 & -0.05 \\
co-added (south)   &  0.20 & -1.68 &  1.54 &  0.78 &  0.16 & -0.04  & -0.05 & 0.48 & -0.04 \\
\enddata
\end{deluxetable}

\subsection{Power asymmetries}

\begin{figure*}
\begin{center}
\includegraphics[width=0.7\linewidth]{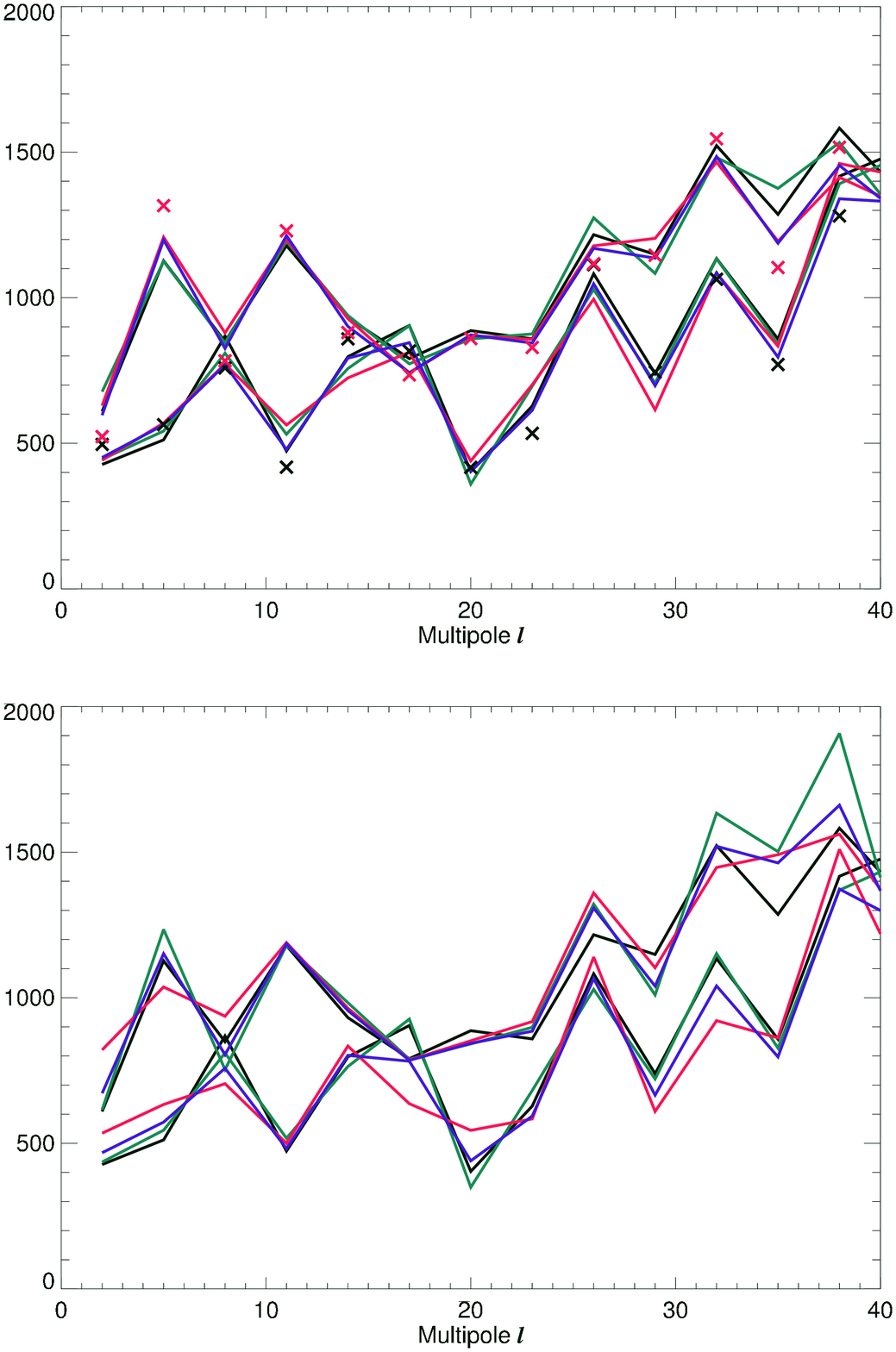}
\caption{ The power spectra taken in the northern (lower lines in each
plot) and southern (higher lines in each plot) hemispheres in the
frame of reference of maximum asymmetry with the north pole at
$(\theta,\phi)=(80^\circ,57^\circ)$ in galactic co-latitude and
longitude. Solid black lines show the results on the co-added ETF map
obtained by the \emph{WMAP} team. Green line shows the WI-FIT result
for the Q band, blue line for the V band, red line for the W band and
black crosses for the Ka band. In the upper plot, the results were
obtained by applying the WI-FIT fitting procedure on the full sky
(outside Kp2). In the lower plot, the fitting procedure was applied
separately to the northern hemisphere for the northern spectra and to
the southern hemisphere for the southern
spectra. \label{fig:cl_hemis}}
\end{center}
\end{figure*}

\citet{asymm1} and \citet{asymm2} reported a significant asymmetry in
the distribution of power in the \emph{WMAP} data. In the reference
frame defined by the north pole at $(\theta,\phi)=(80^\circ,57^\circ)$
(Galactic co-latitude and longitude), the southern hemisphere was
found to have significantly more power on large scales ($\ell<40$)
than the northern hemisphere. We therefore re-estimated the power
spectra independently in these two hemispheres with the WI-FIT
corrected maps, and found remarkably small differences with respect to
the previous analyses (see Fig. \ref{fig:cl_hemis}).  That these two
different approaches to foreground subtraction yield consistent
results strengthens the case against an entirely foreground-based
explanation of the asymmetry.

\begin{figure*}
\begin{center}
\includegraphics[width=0.8\linewidth]{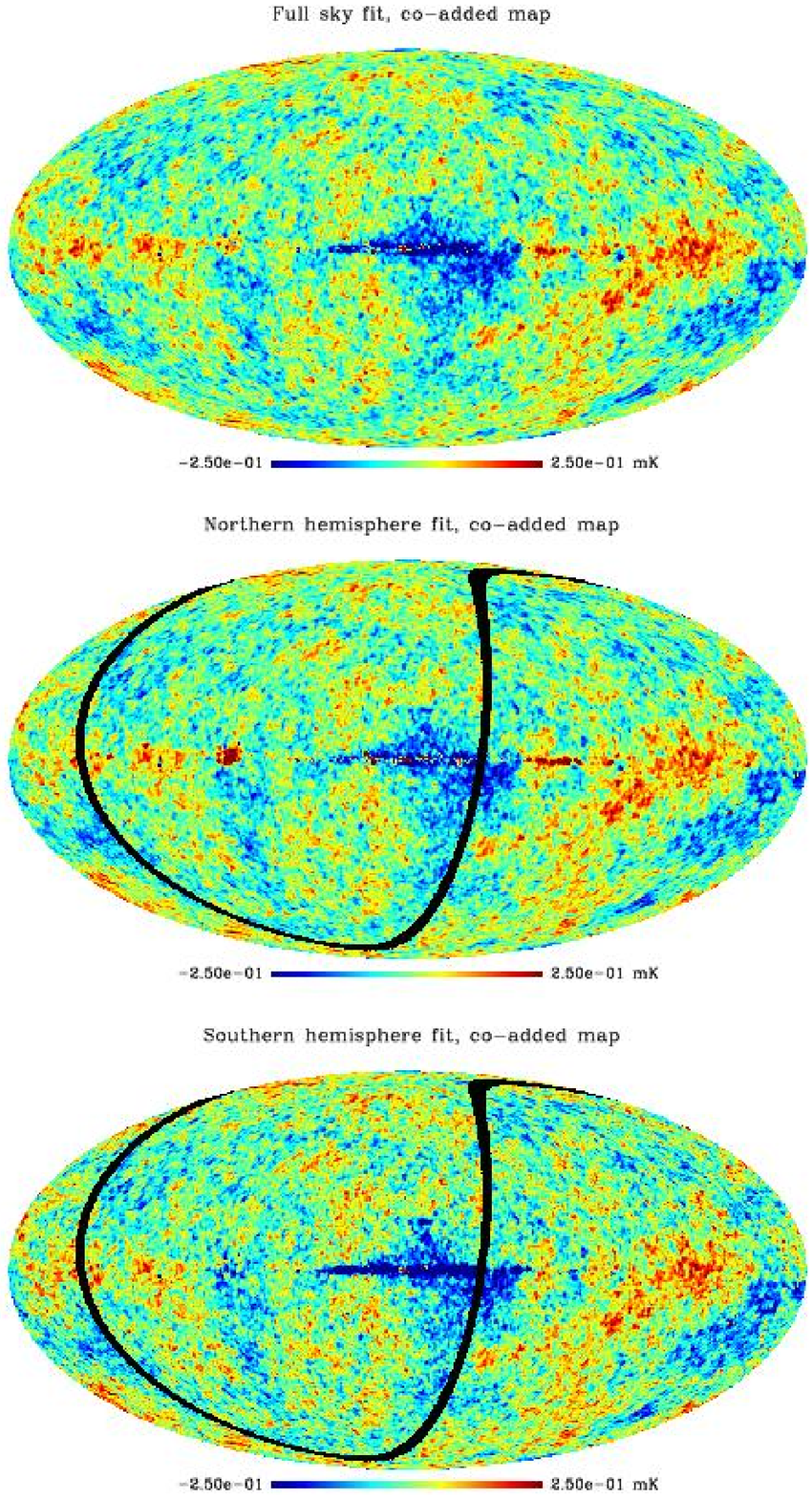}
\caption{ WI-FIT cleaned \emph{WMAP} maps: the maps have been produced
by a noise-weighted combination of the separately cleaned Q, V and W
bands. Upper plot shows the result of the fitting procedure applied to
the full sky (outside the Kp2 cut), middle/lower plot shows the same
map with fitting applied only to the northern/southern (left/right
hemisphere on the plot separated by a black band) hemisphere of maximum
asymmetry.  \label{fig:coaddedmaps}}
\end{center}
\end{figure*}

\begin{figure*}
\begin{center}
\includegraphics[width=0.8\linewidth]{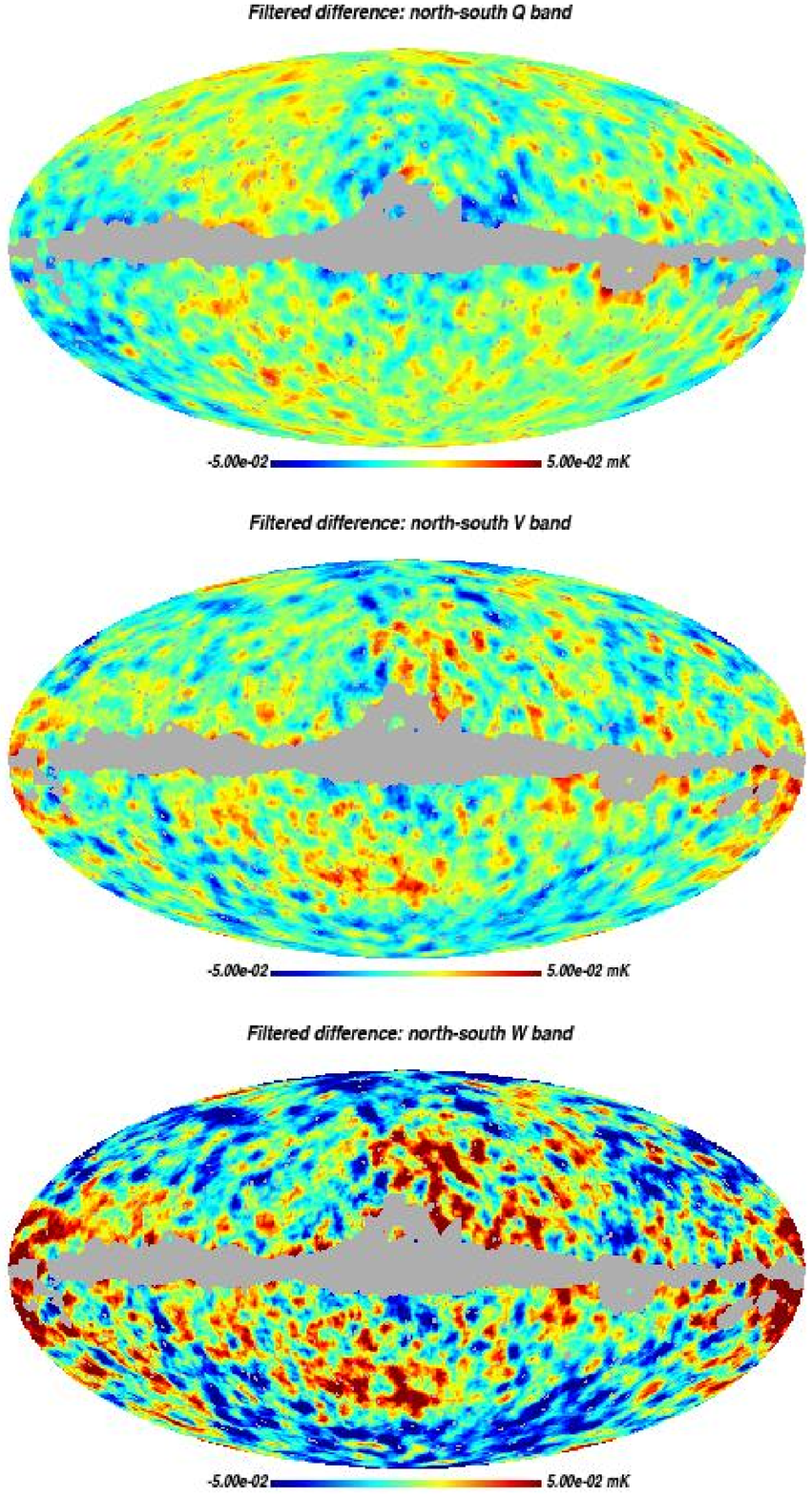}
\caption{ Median filtered (3 degree radius top-hat) differences between the
noise-weighted maps obtained by the best fit template coefficients in
the north and in the south. \label{fig:ns}}
\end{center}
\end{figure*}

However, one flaw in this argument is that the spectral properties of
the foregrounds are assumed to behave uniformly over the sky.  We have
therefore investigated this issue further by applying the WI-FIT
procedure separately to the northern and southern hemispheres.  This
results in two sets of foreground coefficients, one for each
hemisphere, that are subsequently used to generate two different
foreground cleaned maps. These are shown in Fig. \ref{fig:coaddedmaps}
and the difference channel by channel is shown in
Fig. \ref{fig:ns}. We see that the deviations between the maps can be
quite large, even outside the Kp2 cut, supporting the importance of
taking into account spatial variations of the spectral index (see the
discussion in \S \ref{sect:concl}). However, when comparing the
northern hemisphere power spectrum corrected by the northern
coefficients with the corresponding southern hemisphere corrected by
the southern fit, it is evident that a strong asymmetry persists.
(lower plot of Fig. \ref{fig:cl_hemis}).

As an additional check, we have applied the ILC method \citep{ilc3} to
the northern and southern hemisphere separately and derived consistent
results.  Finally, we evaluated the power spectrum on the northern and
southern hemisphere of the map obtained by taking the frequency
combination (2.65Ka-K)/1.65, for which synchrotron emission is
strongly suppressed. Again, the asymmetry was very similar.
If the asymmetry arises as a consequence of residual foregrounds, one
would expect it to vary with the foreground subtraction procedure
applied. The fact that it does not is a strong argument against a
galactic origin for the asymmetry.

\section{Discussion and Conclusions}
\label{sect:concl}

We have introduced a new foreground subtraction method (WI-FIT;
Wavelet based hIghresolution Fitting of Internal templates) that uses
linear combinations of the foreground components, obtained from the
microwave data itself, to fit and subtract foregrounds from the CMB
dominated channels. For high resolution maps, the fitting procedure is
performed in wavelet space since the pixel space approach is
unfeasible for large numbers of pixels. The advantage of this method
is that it relies neither on any assumptions nor prior knowledge about
the galaxy, nor external observations. All foreground information is
obtained from the data themselves by computing differences between
maps at different frequencies.

We have demonstrated that the procedure works well on simulated data,
based on the experimental parameters of the \emph{WMAP} satellite.  We
then applied the method to the \emph{WMAP} data and obtained a
large-scale power spectrum in excellent agreement with that published
by the \emph{WMAP} team (see Fig. \ref{fig:cl_wmap}). We thus obtained
the same results without strong constraints on the galactic emission
based on external templates.  Such a blind analysis method will be of
utmost importance for future polarization experiments, for which
reliable external templates of polarized galactic components may not
be available.

We further showed that the reported power spectrum asymmetry
\citep{asymm1,asymm2} remains unaltered by this approach to foreground
cleaning. This holds true even when the foreground subtraction
procedure is applied in the two opposite hemispheres separately.
We conclude that the asymmetry is seen at the same level in all
frequency bands from $23$ GHz to $94$ GHz and for at least three
different foreground subtraction methods. We thus consider the
possibility of a galactic origin of the asymmetry as remote.


In the present work, we have assumed that the frequency spectra of the
foreground components are independent of position on the sky. This is
not a very realistic assumption and also not a necessary one.  The
WI-FIT method can easily be applied to local regions on the sky
individually, although this will likely introduce a more inhomegeneous
noise structure to the data.
The noisier areas will also need to use more large scale information
in order to obtain a sufficiently high signal-to-noise ratio to avoid
biased template coefficients, as discussed in Appendix \ref{sect:app}.
The wavelet approach presented is well suited suited for such a
scale-dependent procedure.

Finally, we note that the the methods presented here can trivially be
extended to polarization maps of Stoke's parameters Q and U. These
issues will be investigated in a future paper.

\begin{acknowledgments}
  We acknowledge the use of the HEALPix \citep{healpix} package. FKH
  acknowledges support from a European Union Marie Curie reintegration
  grant.  We acknowledge the use of the Legacy Archive for Microwave
  Background Data Analysis (LAMBDA). Support for LAMBDA is provided by
  the NASA Office of Space Science.
\end{acknowledgments}

\begin{appendix}
\section{Procedures for noise bias correction and estimation of residual biases}
\label{sect:app}

Because of the presence of noise in the internal templates, the
template fitting procedure gives biased estimates of the template
coefficients. This is easy to see if we take, as an illustration,
fitting in pixel space of one noisy template in one
channel. Minimizing the $\chi^2$ in equation (\ref{eq:chi2}), we
obtain a best estimate
\begin{equation}
\hat c=\frac{\sum_{ij}\hat s_iC_{ij}^{-1}T_j}{\sum_{ij}\hat s_iC_{ij}^{-1}\hat s_j}.
\end{equation}
Now we replace the external templates by the noisy internal templates
$\hat s_i\rightarrow D_i$ and write the internal templates in terms of
a signal part and a noise part as $D_i=d_i+\delta n_i$. We thus obtain
\begin{equation}
\label{eq:est}
\hat c=\frac{\sum_{ij}(d_i+\delta n_i)C_{ij}^{-1}T_j}{\sum_{ij}(d_i+\delta n_i)C_{ij}^{-1}(d_j+\delta n_j)}=\frac{\sum_{ij}(d_i+\delta n_i)C_{ij}^{-1}(T_j^\mathrm{CMB}+n_i+cd_i)}{\sum_{ij}(d_i+\delta n_i)C_{ij}^{-1}(d_j+\delta n_j)}.
\end{equation}
Taking the ensemble average, we have
\begin{equation}
\VEV{\hat c}=c\VEV{\frac{\sum_{ij}(d_i+\delta n_i)C_{ij}^{-1}d_j}{\sum_{ij}(d_i+\delta n_i)C_{ij}^{-1}(d_j+\delta n_j)}},
\end{equation}
where we have used that the CMB and the noise $n_i$ in the channel to
be cleaned are both uncorrelated with the noise $\delta n_i$ in the
internal template. We see that when the internal template is
noise-free, $\delta n_i\rightarrow0$, the estimate is unbiased
$\VEV{\hat c}\rightarrow c$.

We will now go to wavelet space where we will show how this bias
factor can be significantly reduced. The bias correction procedure
presented in the following can trivially be extended to real space. We
will again start with the simplified case of one template and one
channel. Then we extend the procedure to the realistic situation and
finally we confirm that it works by showing the results from simulated
maps.

To simplify notation, we will now write the sum over wavelet scales as
\begin{equation}
X*X\equiv\sum_{SS'}X_SC_{SS'}X_{S'}
\end{equation}
Then the best estimate can be written as
\begin{equation}
\hat c=\frac{X*X^\nu}{X*X},
\end{equation}
where $X_S=\sum_i(w_{iS}+w_{iS}^n)^2$ and
$X_S^\nu=\sum_i(w_{iS}+w_{iS}^n)w_{iS}^\nu$ (see equation
\ref{eq:xdef}) where $w_{iS}$, $w_{iS}^n$ and $w_{iS}^\nu$ are the
wavelet transforms of $d_i$, $\delta n_i$ and $T_i^\nu$
respectively. Again, the noise term $w_i^n$ is responsible for the
bias and the first step in the bias correction procedure is to remove
the noise variance term from $X_S$
\begin{equation}
X_S\rightarrow X_S-\VEV{(w_i^n)^2}=w_i^2+2w_iw_i^n+\underbrace{(w_i^n)^2-\VEV{(w_i^n)^2}}_{\equiv\delta w_i^2},
\end{equation}
where $\VEV{(w_i^n)^2}$ is obtained from Monte-Carlo simulations of pure noise.
The ensemble average is then (see equation \ref{eq:est})
\begin{equation}
\label{eq:west}
\VEV{\hat c}=c\VEV{\frac{X*(\sum_iw_i(w_i+w_i^n))}{X*X}}.
\end{equation}
We have noted in simulations that 
\begin{equation}
\label{eq:biascond}
\VEV{\hat c}=\VEV{\frac{X*X^\nu}{X*X}}\approx\frac{\VEV{X*X^\nu}}{\VEV{X*X}}
\end{equation}
within a few percent accuracy. This suggests that we could correct for
the bias if we manage to get the ensemble averages of the numerator
and denominator in equation (\ref{eq:west}) equal:
\begin{equation}
\VEV{\hat c}\approx c\frac{\VEV{X*(\sum_i w_i(w_i+w_i^n))}}{\VEV{X*X}}\equiv c\frac{\VEV{N}}{\VEV{D}}.
\end{equation}
The numerator and denominator can be written as
\begin{equation}
\VEV{N}=(\sum_i w_i^2)*(\sum_i w_i^2)+2\VEV{(\sum_i w_iw_i^n)*(\sum_i w_iw_i^n)}
\end{equation}
\begin{equation}
\VEV{D}=(\sum_i w_i^2)*(\sum_i w_i^2)+4\VEV{(\sum_i w_iw_i^n)*(\sum_i w_iw_i^n)}+\VEV{(\sum_i \delta w_i^2)*(\sum_i \delta w_i^2)}
\end{equation}
Clearly, by subtracting the following terms from the denominator, the
average value of the numerator and denominator will become equal and
the estimate $\hat c$ will become unbiased provided equation
(\ref{eq:biascond}) holds. The second step of the bias correction will
thus be
\begin{equation}
\label{eq:step2}
D\rightarrow D-2\VEV{(\sum_i w_iw_i^n)*(\sum_i w_iw_i^n)}-\VEV{(\sum_i \delta w_i^2)*(\sum_i \delta w_i^2)}
\end{equation}
The last term can be easily obtained from Monte-Carlo simulations of pure noise. The first correction term can be obtained from simulations of noise realizations cross-correlated with the internal template $w_i+w_i^n$ taken from the data. Clearly how successful the bias correction is depends on the condition given in equation (\ref{eq:biascond}) and a check of the remaining level of bias would be of high importance. Looking at expression (\ref{eq:west}) it is clear that the exact expression for the bias is given by
\begin{equation}
\frac{\hat c}{c}=\VEV{\frac{X*(\sum_iw_i(w_i+w_i^n))}{X*X}}.
\end{equation}
The problem of obtaining the bias this way is that the noise-free
template $w_i$ is unknown. We can however make the approximation that
our noisy template is a noise-free foreground and then add noise
realizations using Monte-Carlo simulations
\begin{equation}
\label{eq:biasest}
\hat b=\VEV{\frac{X'*(X^\nu)'}{X'*X'}},
\end{equation}
where $X'=\sum_i(w'_i+w_i^N)^2-2\VEV{(\sum_i w_i^n)*(\sum_i w_i^n)}$,
$(X^\nu)'=w'_i(w'_i+w_i^N)-\VEV{(\sum_i w_i^n)*(\sum_i w_i^n)}$,
$w'_i=w_i+w_i^n$ is the noisy template obtained from the data and
$w_i^N$ are Monte-Carlo noise realizations. To correct for the fact
that the templates $w_i$ and $w'_i$ have a slightly different
structure (because of the noise $w_i^n$), we subtract all
contributions of $w_i^n$ to $\VEV{N}$ and $\VEV{D}$ by
\begin{equation}
N\rightarrow N-4\VEV{(\sum_i w_iw_i^n)* (\sum_i w_iw_i^n)}-2\VEV{(\sum_i w_i^n w_i^N)*(\sum_i w_i^n w_i^N)}-\VEV{(\sum_i \delta w_i^2)*(\sum_i \delta w_i^2)},
\end{equation}
\begin{equation}
D\rightarrow D-4\VEV{(\sum_i w_iw_i^n)* (\sum_i w_iw_i^n)}-4\VEV{(\sum_i w_i^n w_i^N)*(\sum_i w_i^n w_i^N)}-\VEV{(\sum_i \delta w_i^2)*(\sum_i \delta w_i^2)},
\end{equation}
obtained with Monte-Carlo simulations in the same manner as above. We
will later show how the bias obtained in this way compares with the
real bias in simulations.

The bias correction procedure described above can easily be extended
to the case with several templates and channels. Instead of making
corrections to the denominator, corrections are made to the matrix
$M_{tf}$ described in section \S \ref{sect:idea}. Using the same
arguments as above, we find that the corresponding bias correction is
given by
\begin{equation}
M_{tf}\rightarrow M_{tf}-M_{tf}^1-M_{tf}^2-M_{tf}^{nn},
\end{equation}
and the estimate of the remaining bias is obtained in the same manner
as above, but again correcting for the additional noise term by
\begin{equation}
M_{tf}\rightarrow M_{tf}-M_{tf}^{n1}-M_{tf}^{n2}-M_{tf}^{n3}-M_{tf}^{n4}-M_{tf}^1-M_{tf}^2-M_{tf}^3-M_{tf}^4-M_{tf}^{nn},
\end{equation}
and
\begin{equation}
B_f^\nu\rightarrow B_f^\nu-\sum_t(M_{tf}^{n3}+M_{tf}^{n4}+M_{tf}^1+M_{tf}^2+M_{tf}^3+M_{tf}^4+M_{tf}^{nn}),
\end{equation}
where
\begin{eqnarray}
M_{tf}^1&=&\VEV{\sum_T (\sum_i (w_i(T)+w_i^n(T))w_i^N(t))*(\sum_i (w_i(T)+w_i^n(T))w_i^N(f))}\\
M_{tf}^2&=&\VEV{\sum_T (\sum_i (w_i(T)+w_i^n(T))w_i^N(t))*(\sum_i (w_i(f)+w_i^n(f))w_i^N(T))}\\
M_{tf}^3&=&\VEV{\sum_T (\sum_i (w_i(t)+w_i^n(t))w_i^N(T))*(\sum_i (w_i(T)+w_i^n(T))w_i^N(f))}\\
M_{tf}^4&=&\VEV{\sum_T (\sum_i (w_i(t)+w_i^n(t))w_i^N(T))*(\sum_i (w_i(f)+w_i^n(f))w_i^N(T))}\\
M_{tf}^{nn}&=&\VEV{\sum T (\sum_i \delta w_i(t)\delta w_i(T))*(\sum_i \delta w_i(f)\delta w_i(T))}
\end{eqnarray}
The matrices $M_{tf}^{n1}-M_{tf}^{n4}$ are defined similarly to
$M_{tf}^{1}-M_{tf}^{4}$, but with $w_i\rightarrow 0$ and the $w_i^n$
are drawn from an independent Monte-Carlo of noise realizations. Note
that the $B_f^\nu$ used for estimating the bias is constructed in the
same manner as the numerator in equation (\ref{eq:biasest}).

Finally these procedures were tested on \emph{WMAP}-like simulations
(described in previous sections). In table (\ref{tab:bias}) we show
the size of the remaining bias on the template coefficients after
applying the bias correction procedure as well as estimating bias with
the above method. We see that the bias is kept at the level of a few
percent (but may be as large as $15\%$ when considering hemispheres
where the signal to noise is lower). Note also that the estimated bias
agrees well with the real bias and in all cases is a conservative
estimate of the bias. Note that we have not included the results from
the W band as one of the internal templates is completely noise
dominated (for the parameters used in the simulation) and the matrix
$M_{tf}$ becomes unstable.

For the \emph{WMAP} data, we estimated the remaining template
coefficient biases in the full sky fits to be less than $5\%$ relative
to the $1\sigma$ error on the coefficients for all bands. In the
hemisphere fits, we found the remaining bias to be in the range
$5-20\%$ of $1\sigma$. The only exception was for the fit to the
northern hemisphere in the W-band for which the bias was estimated to
be $45\%$ of $1\sigma$. Note that the lowest wavelet scales are the
scales most affected by noise. By exluding the noise dominated scales,
the bias factor can be lowered at the cost of larger error bars. For
the fit to the northern hemisphere in the W-band, we remade the
analysis excluding the 3 first wavelet scales and found that the bias
was reduced to a few percent. However, the results obtained with this
map is fully consistent with the map obtained by using all scales and
we have therefore chosen to include only the results of the latter in
this paper.

\begin{deluxetable}{ccc}
\tabletypesize{\small}
\tablecaption{The real and estimated bias on the internal template coefficients obtained in \emph{WMAP}-like simulations.\label{tab:bias}}
\tablewidth{0pt}
\tablehead{
\colhead{\emph{WMAP} band/region} & \colhead{real bias range} & \colhead{estimated bias range}  }
\startdata

Q  & $\sim0.1\%$ & $\sim 1\%$ \\
V  & $\sim1\%$ & $3-4\%$ \\
Q (north) & $\sim1\%$ & $\sim 1\%$\\
V (north) & $1-4\%$ & $2-6\%$\\
Q (south) & $2-4\%$ & $2-4\%$ \\
V (south) & $10-15\%$ & $10-20\%$ 
\enddata
\end{deluxetable}

\end{appendix}

\end{document}